\documentclass[10pt, conference, compsocconf]{IEEEtran}
\usepackage{stfloats}
\usepackage{graphicx}
\usepackage{tabularx}

\begin{document}
\title{A Critique of Human-Autonomous Team Dynamics: Contrasting Qualitative and Quantitative Perspectives}

\author{
    \IEEEauthorblockN{Hanjing Shi\IEEEauthorrefmark{1}}
    \IEEEauthorblockA{\IEEEauthorrefmark{1}Department of Computer Science and Engineering\\
    Lehigh University\\
    Bethlehem, Pennsylvania\\
    Email: hasa23@lehigh.edu}
}

\maketitle

\begin{abstract}
The critique paper offers a comprehensive analysis of two seminal works The critique paper provides an in-depth analysis of two influential studies in the field of Human-Autonomous Teams (HATs). \textit{Musick et al.} explored qualitative dimensions of HAT dynamics, examining the influence of team composition on emotions, cognitive processes, and the development of team cognition. Their research revealed that teams with a majority of human members, known as Multi-Human HATs, generally surpass Multi-Agent HATs in performance, highlighting the critical influence of human perception on team dynamics. Employing qualitative interview analysis anchored in theoretical frameworks, \textit{Musick et al.} captured the detailed subtleties of participants' experiences. In contrast, \textit{Schelble et al.} utilized a quantitative methodology to provide data-driven insights into how the perception of AI teammates affects team performance. Despite the rich insights from \textit{Musick et al.}'s qualitative research, their findings face limitations in terms of broader applicability. Both \textit{Musick et al.} and \textit{Schelble et al.} agree in their conclusions that Multi-Human HATs typically outperform their Multi-Agent counterparts, again emphasizing the crucial role of human perception in team dynamics. The critique paper suggests that future research should focus on understanding perceptions of teams heavily reliant on AI. Such investigations could illuminate how trust and skepticism are shaped in teams where AI plays a dominant role.

\end{abstract}

\IEEEpeerreviewmaketitle

\section{Introduction}

Recent advances in computational science, artificial intelligence (AI), human-automation interaction (HAI), and human-computer interaction (HCI) have given rise to ``synthetic agents'' serving as integral members of teams. Such a development now stands at the forefront of HCI research, focusing on the intricate interplay between human capabilities and the computational power of AI \cite{chen2018human}. The integration of AI into team-based settings has given rise to Human-Autonomous Teams (HATs), a transformed paradigm characterized by the interdependent collaboration between humans and AI entities. Within HATs, the synergy between human and artificial agents spans a spectrum of activities, from decision-making processes to complex communication patterns, fundamentally altering the traditional dynamics of teamwork \cite{o2018creating} \cite{salas2008teams} \cite{mathieu2018evolution}. The body of research on HATs is expansive, investigating the intricate dynamics and processes that manifest within teams composed of both humans and autonomous systems.

Convergence of human intuitive reasoning and AI's methodical computation introduces a revolutionary collaborative model. Pioneering studies by \textit{Musick et al.} \cite{musick2021happens} and \textit{Schelble et al.}\cite{schelble2023investigating} stand out for their exploration of how team composition perception—specifically the presence of AI agents—affects the internal dynamics of HATs. Their research offers vital insights into the complex interactions and cooperative processes that define human-agent teams\cite{musick2021happens} \cite{schelble2023investigating}. Through their investigative lens, \textit{Musick et al.} and \textit{Schelble et al.} have contributed to a deeper understanding of HATs, examining how the intermingling of human and AI team members influences collective efficiency, cognition, and coordination. As such, their work provides a crucial foundation for future inquiry into the optimization of human-AI integration in collaborative settings, with the goal of harnessing the full potential of these heterogeneous teams.

The primary objective of critique paper is to understand how perceptions of team composition influence sentiments towards teammates, team processes, cognitive states, and the emergence of a system of team cognition. The methodology of the paper contrasts two seminal studies from \textit{Musick et al.} and \textit{Schelble et al.}, emphasizing their distinct strengths, overlapping themes, and variances. Additionally, the critique aims to give a thorough overview of the HAT landscape by combining observations from both qualitative and quantitative methods. The two papers provided investigate similar research questions regarding the effect of perceived AI teammates on team performance and cognition in human-AI teams. However, the papers adopt different research perspectives and methodological approaches. The paper by \textit{Musick et al.} uses qualitative methods to study the influence of perceptions on team composition, distinguishing between perceived Multi-Agent HATs and actual human participants. \textit{Musick et al.}'s approach examines how such perceptions might affect fundamental team processes, including communication, implicit coordination, and shared mental models. Conversely, \textit{Schelble et al.} employ quantitative methods, elucidating the interaction between perceived teammate artificiality, task difficulty, and human task performance. To provide a clearer understanding of their focus, the following research questions were central to their studies:

\noindent \textbf{\textit{Musick et al.}:} \\
\noindent RQ: ``How does the perceived composition of a team affect sentiment towards teammates, team processes, and the emergence of a system of team cognition?''

\noindent \textbf{\textit{Schelble et al.}:} \\
\noindent RQ1: ``How does the perception of working with an AI teammate, as opposed to a human teammate, influence human-task performance based on task difficulty?''

\noindent RQ2: ``How does this perception influence overall team cognition?''

\section{Methodology}

In \textit{Musick et al.}'s study, the focus is drawn towards deciphering how the perceived composition of a team influences various aspects such as sentiments toward team members, cognitive states, communication, and overall team cognition. By adopting an IIHAT (Implicit Interaction for Human-Autonomy Teams) simulation game combined with post-game focus group interviews. The IIHAT task was developed explicitly for human-agent teaming research and did not allow communication between players to isolate the effect of implicit communication on team cognition and human-agent teams \cite{entin1999adaptive} \cite{hanna2014impact}\cite{hanna2015impact}\cite{matessa2017human}.

In Schelble's study, the emphasis was on understanding the nuanced effects of AI inclusion on human performance, especially in relation to task difficulty. The research employed the same IIHAT simulation platform but approached the study from a quantitative perspective. The primary objective was to measure human task performance across different maps and levels of difficulty. Additionally, participants' mental models were elicited using the paired sentence comparison method for its proven reliability in eliciting mental models, as outlined in Mathieu et al. (2000) and others. This method involved participants comparing various task and team-related attributes to determine their relationships as positive, negative, or non-existent, with attributes drawn from detailed task analysis and past literature.

\subsection{IIHAT simulation}

\begin{figure*}[!t]
    \centering
    \includegraphics[width = 10 cm]{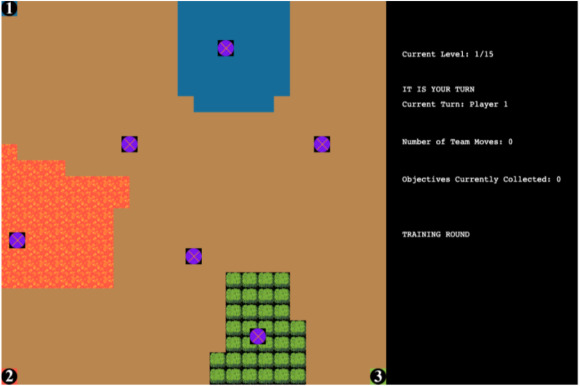}
    \caption{IIHAT task simulation.}
    \label{fig:map}
\end{figure*}

The IIHAT (Implicit Interaction for Human-Autonomy Teams) simulation served as the experimental platform for both studies under critique. During the simulation, participants were tasked with escaping an island, a challenge necessitating distinct abilities to navigate different terrains and demanding seamless teamwork for success. The simulation was designed to prevent verbal or textual communication among players, compelling them to rely on in-game actions and strategies as subtle signals of their intentions and plans. The studies recruited undergraduate participants to explore the dynamics of human interaction within a HAT environment and to deduce underlying themes regarding human perception and trust towards autonomous agents. (Figure ~\ref{fig:map})

\subsection{``Wizard of Oz" (WoZ) Approach}

In both \textit{Musick et al.}'s and \textit{Schelble et al.}'s experiments, the ``Wizard of Oz" (WoZ) agent introduces an intriguing element during the Multi-Agent HAT (Human-Agent Teaming) scenario. This agent, rather than being an actual AI, was a human participant designed to emulate typical human interactions. The WoZ agents were carefully crafted to reflect human behaviors observed in traditional human teams. The primary aim of their projects was to assess participants' perceptions and behaviors under the belief that they were collaborating with an AI entity. This approach was crucial not only for studying human responses to AI without the complexities or unpredictability of genuine AI behavior but also for uncovering perceptions and behaviors when participants are under the assumption that they are interacting with AI.

\subsection{Qualtative Methods in \textit{Musick et al.}'s Study}

In the \textit{Musick et al.}'s study under critique, a meticulous experimental design was employed to understand the dynamics of Human-Autonomous Teams (HATs). One hundred undergraduate participants, with a gender distribution of 67\% female, were grouped into forty-six teams. Within the study, teams were organized into three distinct types, each comprised of three members: Traditional Human Teams, Multi-Human HATs, and Multi-Agent HATs. Traditional Human Teams were composed entirely of human members. In the Multi-Human HATs, the teams included two human members and one autonomous agent. Conversely, Multi-Agent HATs were formed with one human member and two autonomous agents. The study featured 15 teams in both the Traditional Human Teams and Multi-Human HATs categories, and 16 teams in the Multi-Agent HATs category. Additionally, an agent called the WoZ was presented in the Multi-Agent HAT scenario.  The WoZ agent was not an AI, but rather a human player whose behavior was crafted to mimic typical human interactions.

A manipulation check, a standard procedure in experimental research, was implemented to confirm whether participants’ perceptions aligned with the intended experimental conditions. Specifically, researchers needed to verify that participants believed they were interacting with artificial intelligence during the study. Conducted subsequent to the focus group discussions, the manipulation check demonstrated a high level of effectiveness. Out of the total participant pool, only four individuals—two from Multi-Human Human-Autonomous Teams and two from Multi-Agent Human-Autonomous Teams—expressing doubts about interacting with AI. Such results indicate that the experiment successfully conveyed the impression of AI interaction to the majority of participants. 

\subsection{Quantitative Methods in \textit{Schelble et al.}'s Study}

Conversely, \textit{Schelble et al.}'s study took a quantitative route, utilizing the same IIHAT simulation to assess the dynamics of Human-Autonomous Teams. The study's experimental design included the Wizard of Oz methodology to simulate an AI agent's presence and measured various aspects such as task performance, mental model development, and team cognition. \textit{Schelble et al.} recruited 75 undergraduate participants, forming 30 teams. Half of these teams (15) participated in each of the two conditions: Human-Human-Human (HHH) and Human-Human-Agent (HHA). The quantitative data provided empirical evidence on the effects of perceived teammate artificiality and task difficulty on human performance within HATs.

In \textit{Schelble et al.}'s study, a comprehensive set of measures was utilized to evaluate various aspects of the experiment involving one hundred undergraduate participants. The study structured the teams into two main categories: Traditional Human Teams (HHH) and Human-Human-Agent Teams (HHA). Each team, regardless of type, comprised three members. In the HHH teams, all three members were humans, while in the HHA teams, two human members were joined by an autonomous agent simulated by a WoZ confederate.

In the research by \textit{Schelble et al.}, individual human task performance within teams was evaluated by averaging the actions of all human team members, while specifically excluding the moves made by the WoZ confederate that represented the autonomous agent in HHA teams. The similarity of mental models within these teams was effectively measured using the Pathfinder network-scaling algorithm, a widely recognized tool in shared mental model research. This algorithm converts participants' pairwise comparisons into graphical representations of mental model networks to compute a similarity score for each team dyad, ranging from 0 (indicating no similarity) to 1 (denoting perfect similarity). The use of the Pathfinder network-scaling algorithm was pivotal in shedding light on the extent to which team members shared and aligned their cognitive frameworks.

The perception of team cognition, an essential aspect of team dynamics, was evaluated using the Teamwork Schema Questionnaire, which helped gauge how team members viewed the collective cognitive functioning and efficiency of their team. Additionally, to understand participants' general acceptance and attitudes towards artificial intelligence, a modified version of the UTAUT (Unified Theory of Acceptance and Use of Technology)\cite{pontiggia2010network} was employed by \textit{Schelble et al.}. In their adaptation, the term 'technology' was specifically replaced with 'AI' to tailor the questionnaire to the context of the study, thereby providing insights into the participants' receptiveness to AI in team settings.

Moreover, \textit{Schelble et al.}'s research utilized an experimental approach with two conditions of perceived teammate artificiality across three stages of task difficulty. Each condition completed fifteen rounds of a human-AI task simulation known as ``IIHAT'' (Implicit Interaction for Human-Autonomy Teams) \cite{musick2021happens}. The quantitative findings showcased the effects of perceived teammate artificiality and its interaction with task difficulty on human task performance. Additionally, the research highlighted the effect of perceived teammate artificiality on the development of shared mental models and perceived team cognition.

In essence, \textit{Schelble et al.}'s quantitative approach provided a structured and empirical lens to understand the nuances of human-agent interaction within a team setting. The meticulous design and measures ensured a thorough examination of the participants' perceptions, behaviors, and performance in the context of HATs.

\subsection{Comparative Analysis}

Both studies, while utilizing the same IIHAT simulation platform, offered distinct methodologies and insights. The comparative analysis of these two studies highlights the distinct contributions of qualitative and quantitative research methodologies to the understanding of HATs.  \textit{Musick et al.}'s qualitative analysis offered rich, narrative insights into participant perceptions, while \textit{Schelble et al.}'s quantitative analysis provided structured empirical data on performance metrics. 

Both experiments conducted a comprehensive analysis of participant interactions within a HAT environment, meticulously collecting and analyzing data to uncover themes related to human perception and trust towards autonomous agents. By providing a holistic understanding of the dynamics within Human-Autonomy Teams (HATs), the studies by \textit{Musick et al.} and \textit{Schelble et al.} underscore the importance of integrating diverse research approaches to fully comprehend the HCI field and offer a comprehensive view of the complex dynamics at play in HAT interactions.

\section{Discussion}

\subsection{Qualtative Results in \textit{Musick et al.}'s Study}

\begin{figure*}[!t]
\centering
\includegraphics[width = 11 cm]{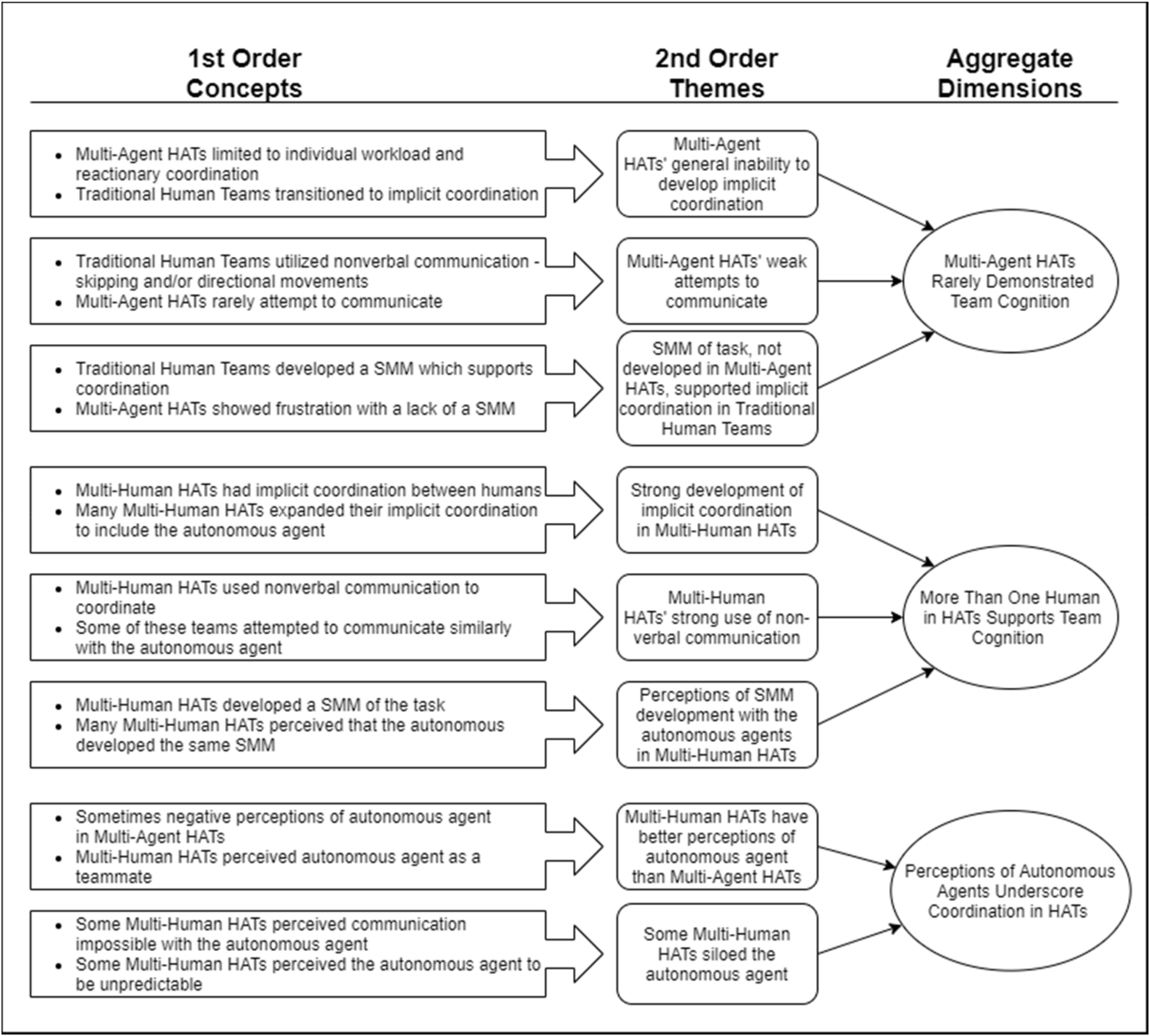}
\caption{Data structure of thematic analysis.}
\label{fig:graph}
\end{figure*}

The thematic analysis extracted from the figure offers profound insights into the dynamics of different team compositions, focusing on communication, coordination, and perceptions.

\subsubsection{Traditional Human Teams Dynamics}

Contrasting with Multi-Agent Human-Autonomy Teams, Traditional Human Teams demonstrate a refined level of coordination, primarily implicit, facilitated by non-verbal cues and signals that indicate a strong rapport among team members. Non-verbal coordination among humans typically involves implicit mechanisms, including the use of timing, movement patterns, and other behavioral cues that do not rely on spoken or written communication. For example, in a task-based scenario, team members might coordinate their actions based on the observed actions of others, timing their responses or movements in sync with their teammates. The development of Shared Mental Models (SMM) within these teams enhances task coordination, allowing for anticipation, support, and synergy in actions that boost operational efficiency.

\subsubsection{Multi-Agent HATs Dynamics}

Teams characterized by the presence of multiple artificial intelligence agents, known as Multi-Agent HATs, appear to operate within an individualistic framework. They concentrate on single tasks, reflecting a design or a nature tailored to autonomous functioning rather than collaborative effort. Coordination within these teams is reactive, with limited proactive strategy formulation. A critical observation is their minimal communication efforts, which, given the centrality of communication in teamwork, may severely limit their cohesion and overall effectiveness.

\subsubsection{Multi-Human HATs Dynamics}

Multi-Human HATs, bridging the dynamics between the former two categories, exhibit both the implicit coordination found in human teams and the challenge of integrating AI agents. They utilize non-verbal communication to coordinate not only among humans but also in attempts to include autonomous agents, displaying both innovation and complexity in strategy integration. The perceptions of autonomous agents within these teams are varied, ranging from unpredictability to acceptance as team members. These perceptions highlight the challenges of integration and suggest a need for more precise integration strategies, which could involve modifications in AI design or advancements in team training to promote positive and consistent team interactions.

From a broader perspective, the figure suggests that successful interactions in HATs are based on effective communication, a shared understanding of tasks, and the adaptability of members, especially concerning autonomous agents. To advance in the realm of HATs, future research and development should focus on enhancing these fundamental aspects, aiming to create more harmonious and efficient teams.(Figure ~\ref{fig:graph})

\subsection{Quantitative Results in \textit{Schelble et al.}'s Study}

The study conducted by \textit{Schelble et al.} was designed to investigate the effects of AI teammates on the collective performance of human-AI teams and to examine the perceptions human team members developed about their non-human counterparts. The results revealed a distinct impact of perceived teammate identity on performance: individuals who believed their third teammate was AI exhibited lower performance compared to those who believed their teammate was human. However, their result was not uniform across all levels of task difficulty. 

Intriguingly, when faced with the most challenging tasks, the groups working with perceived AI teammates demonstrated a significant improvement in performance, surpassing the groups with perceived human teammates. \textit{Schelble et al.}'s outcome stands in contrast to the traditional findings within the human-AI teaming literature, which often do not show such a positive impact of AI perception in high-difficulty situations. The research by \textit{Schelble et al.} thus adds a new dimension to our understanding of human-AI interaction, highlighting how the presence of AI can be particularly advantageous in demanding and complex task environments.

\begin{table}[ht]
\centering
\caption{Mean and standard deviations for dependent variables.}
\label{tab:my-label}
\small
\begin{tabularx}{\columnwidth}{l|l|l|l|l}
 & \multicolumn{2}{c|}{HHH} & \multicolumn{2}{c}{HHA} \\
Measure & Mean (N) & SD & Mean (N) & SD \\
\hline
Total Human & 357.73 (15) & 4.70 & 362.93 (15) & 5.05 \\
Easy Map & 106.27 (15) & 2.30 & 110.90 (15) & 1.40 \\
Medium Map & 118.84 (15) & 2.75 & 124.43 (15) & 3.70 \\
Hard Map & 132.98 (15) & 3.30 & 127.80 (15) & 2.08 \\
Team Cognition & 7.02 (15) & 3.07 & 10.13 (15) & 4.77 \\
Team Mental Model & 0.31 (15) & 0.09 & 0.26 (15) & 0.09 \\
Task Mental Model & 0.50 (15) & 0.13 & 0.52 (15) & 0.13 \\
\end{tabularx}
\end{table}

Furthermore, in \textit{Schelble et al.}’s investigation into the effects of perceived teammate artificiality on human performance and cognition by providing a multifaceted view of human-AI team dynamics. Firstly, the total task performance of teams with an AI agent (HHA) was slightly higher (Mean = 362.93, SD = 5.05) compared to all-human teams (HHH) (Mean = 357.73, SD = 4.70). The trend was particularly evident in tasks of varying difficulties. For easy and medium difficulty tasks, HHH teams demonstrated superior performance, with means of 106.27 (SD = 2.30) and 118.84 (SD = 2.75) respectively. In contrast, HHA teams outperformed HHH teams in hard difficulty tasks (Mean = 127.80, SD = 2.08 for HHA vs. 132.98, SD = 3.30 for HHH). 

Additionally, \textit{Schelble et al.}’s study examined perceptions of team cognition, revealing that participants in HHH teams perceived better team cognition (Mean = 7.02, SD = 3.07) compared to those in HHA teams (Mean = 10.13, SD = 4.77), suggesting a negative impact of AI presence on perceived team efficiency and synergy. Furthermore, the mental model similarity between team members was found to be slightly better in human-only teams for team mental models (Mean = 0.31, SD = 0.09 for HHH vs. 0.26, SD = 0.09 for HHA), while task mental model similarity was marginally higher in HHA teams, though not significantly different. (Table~\ref{tab:my-label})

\begin{figure*}[!t]
    \centering
    \includegraphics[width = 12 cm]{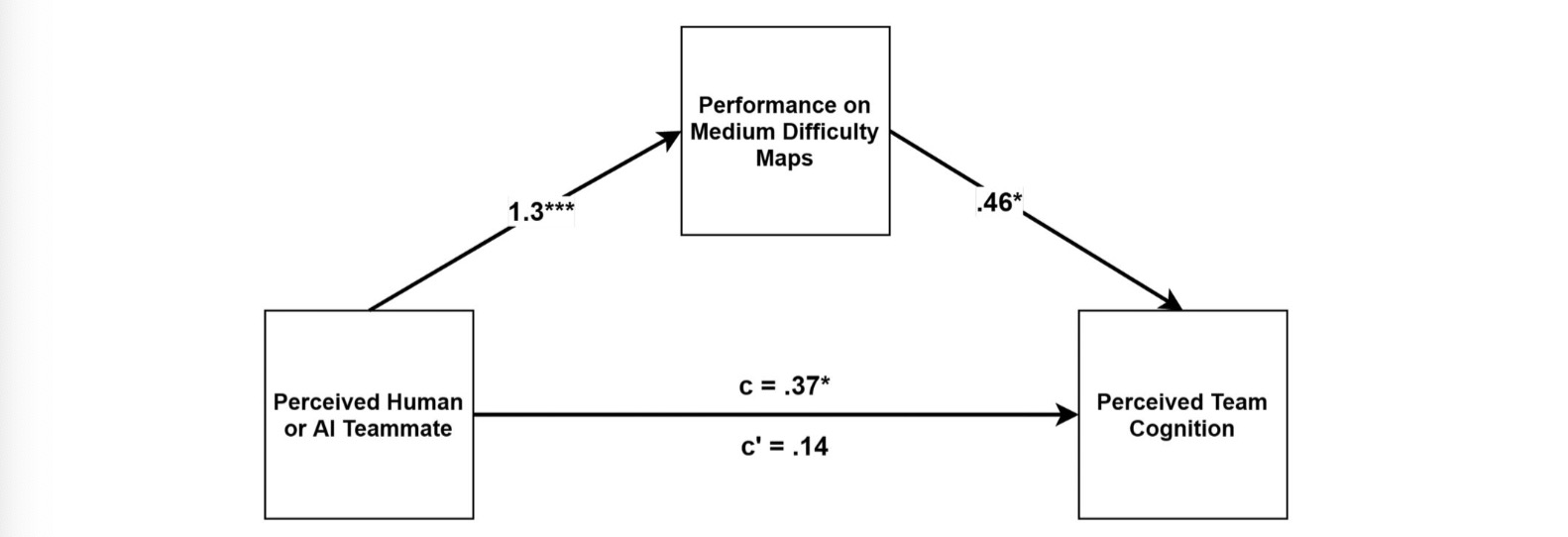}
    \caption{Mediating effect of medium map.}
    \label{fig:graph3}
\end{figure*}

The study by \textit{Schelble et al.} presents a detailed examination of how perceived AI teammates affect human task performance and team cognition. The nuanced findings, presented in a mediation model, suggest a slight overall advantage in task performance for Human-Human-Agent (HHA) teams on medium difficulty maps. The model illustrates that AI teammates might enhance task efficiency, particularly in tasks requiring increased cognitive engagement. It reveals that performance enhancement serves as a mediating factor, where the perception of an AI teammate influences performance outcomes, which in turn affect team cognition.

The findings of \textit{Schelble et al.}'s study underscore the significance of task performance, particularly on medium difficulty tasks, in shaping the influence of perceived teammate artificiality on team cognition. Specifically, variations in performance were notably affected by the perception of a teammate as AI, as shown by a significant path coefficient of 1.3. The study reveals that the impact of perceived teammate artificiality on team cognition is primarily indirect, mediated through performance on medium difficulty tasks (indicated by a path coefficient of .46). Although the overall effect of perceived artificiality on team cognition is significant (c = .37), its direct influence (c' = .14) becomes non-significant when accounting for the mediating effect of task performance. The statistical results from \textit{Schelble et al.}'s research suggest that the perceived artificiality of a teammate on team cognition is fully mediated by participant performance on tasks of medium difficulty, according to Schelble et al.'s research. (Figure ~\ref{fig:graph3})

Despite the nuanced influence of perceived AI on team cognition through task performance, the challenge persists in seamlessly integrating AI agents into the established cognitive framework typically seen in human teamwork. The data indicate that AI teammates can inadvertently affect the team's perceived unity and shared cognitive function, likely stemming from divergent communication styles and coordination methods between human members and AI agents. The research by \textit{Schelble et al.} on mental model similarities provides empirical evidence supporting this point: teams composed solely of humans exhibit a higher degree of synchrony in their team dynamics. In contrast, teams with both human and AI agents, referred to as Human-Human-Agent (HHA) teams, demonstrate a slight advantage in cognition that is directly related to the completion of tasks.

In summary, Schelble et al.'s research brings to light the intricate dynamics of human-AI interaction, highlighting the capacity of AI agents to enhance task performance in complex tasks through their computational capabilities. These findings offer substantial insights into the dynamics of human-AI teams, indicating that the perception of AI teammates indirectly affects team cognition by altering human performance on tasks that require a medium level of difficulty. The insights garnered from \textit{Schelble et al.} research lay the groundwork for further exploration into the development and implementation of AI within team settings, emphasizing the need for a nuanced approach to the integration of AI agents into human teams.

\section{Limitations}

The research conducted by \textit{Musick et al.}, which was pioneering in its examination of three-member Human-Autonomy Teams (HATs) in action-oriented tasks, encounters specific limitations due to its highly controlled environment. The focus on short-duration tasks with limited communication raises questions about the generalizability of their findings to more diverse and realistic HAT scenarios, particularly those involving longer interactions or varied team compositions. The methodological choices of \textit{Musick et al.}, especially the constraints on communication and team size, may have limited the depth of team dynamics observed, potentially omitting richer interactions common in more open and naturalistic settings. Their findings also underscore the need for future research into the design and perception of autonomous agents as integral team members, emphasizing the potential benefits of cross-training and empathy-building interventions. Additionally, their work suggests a critical need for longitudinal research in HATs to understand how these dynamics evolve over time, particularly in relation to the maturation process in teams and the utilization of each member's strengths.

\textit{Schelble et al.}'s study, while concentrating on the effects of perceived teammate artificiality, faces its own limitations due to the non-communication protocol among participants and the simplicity of the task employed. The methodological decisions of \textit{Schelble et al.}, crucial for isolating certain variables, may limit the practical application of their findings, potentially not reflecting the complex dynamics of human-AI interactions in real-world settings. Their study's approach does not fully consider how individual and group interactions with AI might evolve in environments that facilitate diverse forms of communication and task complexity. Consequently, future research should extend beyond the current study’s scope to encompass a variety of tasks and settings, particularly exploring how different communication strategies impact human-AI collaboration in practical contexts.

When considered together, the research conducted by both \textit{Musick et al.} and \textit{Schelble et al.} utilizes the HATII task. Although their approach provides a standardized method to measure human-AI interaction, it may not accurately capture the diversity and complexity of real-world scenarios where HAT dynamics are more pronounced. By focusing on a specific, controlled task, both studies potentially overlook the influence of various task types, especially those with higher variability and unpredictability, on human-AI team dynamics.

Additionally, the use of the WoZ methodology, where human operators simulate autonomous agents, presents concerns regarding the authenticity of the AI's behavior and decision-making processes. Since the simulation may not fully represent the capabilities and limitations of actual autonomous agents, particularly in terms of autonomy, learning ability, and adaptability, the findings from both \textit{Musick et al.} and \textit{Schelble et al.}'s studies might not be entirely generalizable to scenarios involving real AI systems.

These limitations underscore the need for employing a broader range of tasks in future research, including those that more closely mirror real-world challenges and contexts, to better understand the versatility and adaptability of HATs in various settings. Future studies should aim to incorporate actual AI agents with varying levels of autonomy and complexity to provide a more accurate depiction of human interaction with AI.

\section{Future Works}

The future of Human-Autonomy Teams (HATs) lies in addressing key challenges, focusing on the design and perception of autonomous agents, and understanding the long-term evolution of human-AI relationships across diverse settings. A critical step is to assess how the findings of \textit{Musick et al.} and \textit{Schelble et al.}'s studies can be applied to real-world human-AI team interactions, offering valuable insights. A particularly unexplored area for future exploration is the nuanced impact of Generative AI on team dynamics, interpersonal relations, and collective intelligence within team-based work environments. Although AI-driven decision-making and task automation have the potential to optimize team efficiency, they also introduce challenges such as unclear attribution of contributions, biases in AI algorithms, and the risk of dehumanizing work experiences\cite{westby2023collective}. Future studies aim to reveal how the less visible roles of Generative AI tools as behind-the-scenes assistants affect team interactions and outcomes, focusing particularly on trust dynamics and contribution considerations, which are pivotal in shaping collective intelligence\cite{dwivedi2021artificial}.

As AI becomes more prevalent in team-based work environments and collaborative projects, there is a growing need to enhance mutual understanding, effective communication, and adaptability between human and AI team members. To address the limitations identified in the critiques of previous studies, future research will expand our understanding of human-AI team interactions by examining AI's role as an unseen teammate and its influence on group dynamics. The methodology will encompass sociological approaches and an iterative design process. Tools like the Moshi platform, which currently aids in testing smart reply techniques in conversational pairs, will be developed further to support larger teams and integrate leading Generative AI technologies\cite{hohenstein2021artificial}. This advancement will facilitate the study of AI-mediated communication within teams. Future research will focus on investigating the impact of the visibility of AI assistance on team members’ decision-making, trust dynamics, and engagement with AI tools, thereby influencing collective intelligence.

Although previous work highlights significant gaps, potential challenges, and opportunities for innovation in human-autonomy team dynamics, ongoing research in human-AI collaboration is expected to lead to the development of more integrated and efficient team dynamics. The effectiveness of future human-AI joint endeavors will rely on a harmonious blend of AI's computational prowess with human creativity and problem-solving skills.

\section{Conclusions }

The critique combines significant insights from two foundational studies by \textit{Musick et al.} and \textit{Schelble et al.} to enhance our understanding of the complex dynamics in Human-Autonomy Teams (HATs). The research by \textit{Musick et al.} reveals the inherent challenges in teams comprising multiple autonomous agents. Notably, their findings indicate that Multi-Agent HATs tend to be less effective than teams predominantly consisting of human members, a situation arising from the social cognitive challenges involved in recognizing autonomous agents as full team members. \textit{Musick et al.} highlight the criticality of designing autonomous agents in a way that fosters their perception as independent, intelligent entities with volitional capabilities. Their research underscores the necessity for ongoing, long-term studies in the field of HATs, suggesting that the interplay between humans and autonomous agents is likely to evolve, enabling these teams to progressively optimize the use of each member's unique strengths.

On the other hand, \textit{Schelble et al.}'s study underscores the nuanced impacts of perceived AI teammates on human performance and cognition in HATs. Their findings revealed that the presence of AI agents can enhance task performance in challenging scenarios, but it also negatively influences perceived team cognition. \textit{Schelble et al.}'s result suggests a complex interplay between human perception of AI teammates and team dynamics, where AI agents, despite their computational capabilities, struggle to seamlessly integrate into the human cognitive framework of teamwork.

In conclusion, the critique merges insights from \textit{Musick et al.} and \textit{Schelble et al.}'s studies, highlighting a transitional phase in the evolution of team dynamics, moving from purely human teams to those increasingly influenced by AI. While AI agents can augment task-specific performance, their integration into team cognition and coordination remains a significant challenge. The insights from \textit{Musick et al.}  and \textit{Schelble et al.} contribute substantially to our understanding of HAT dynamics, offering a foundation for further exploration in the development and implementation of AI in team settings.

\section{Acknowledgment}

The authors would like to thank Dr.Dominic DiFranzo for his invaluable advice and guidance throughout this research.

\bibliographystyle{abbrv}
\bibliography{ref}

\begin{thebibliography}{10}

\bibitem{chen2018human}
J.~Y. Chen.
\newblock Human-autonomy teaming in military settings.
\newblock {\em Theoretical issues in ergonomics science}, 19(3):255--258, 2018.

\bibitem{dwivedi2021artificial}
Y.~K. Dwivedi, L.~Hughes, E.~Ismagilova, G.~Aarts, C.~Coombs, T.~Crick, Y.~Duan, R.~Dwivedi, J.~Edwards, A.~Eirug, et~al.
\newblock Artificial intelligence (ai): Multidisciplinary perspectives on emerging challenges, opportunities, and agenda for research, practice and policy.
\newblock {\em International Journal of Information Management}, 57:101994, 2021.

\bibitem{entin1999adaptive}
E.~E. Entin and D.~Serfaty.
\newblock Adaptive team coordination.
\newblock {\em Human factors}, 41(2):312--325, 1999.

\bibitem{hanna2014impact}
N.~Hanna and D.~Richards.
\newblock The impact of communication on a human-agent shared mental model and team performance.
\newblock In {\em Proceedings of the 2014 international conference on Autonomous agents and multi-agent systems}, pages 1485--1486, 2014.

\bibitem{hanna2015impact}
N.~Hanna, D.~Richards, et~al.
\newblock The impact of virtual agent personality on a shared mental model with humans during collaboration.
\newblock In {\em Aamas}, pages 1777--1778, 2015.

\bibitem{hohenstein2021artificial}
J.~Hohenstein, D.~DiFranzo, R.~Kizilcec, Z.~Aghajari, H.~Mieczkowski, K.~Levy, M.~Naaman, J.~Hancock, and M.~Jung.
\newblock Artificial intelligence in communication impacts language and social relationships. arxiv.
\newblock {\em arXiv preprint arXiv:2102.05756}, 2021.

\bibitem{matessa2017human}
M.~Matessa, V.~Battiste, and W.~W. Johnson.
\newblock Why human-autonomy teaming.
\newblock In {\em Advances in Neuroergonomics and Cognitive Engineering: Proceedings of the AHFE 2017 International Conference on Neuroergonomics and Cognitive Engineering, July 17--21, 2017, The Westin Bonaventure Hotel, Los Angeles, California, USA}, volume 586, page~3. Springer, 2017.

\bibitem{mathieu2018evolution}
J.~E. Mathieu, M.~A. Wolfson, and S.~Park.
\newblock The evolution of work team research since hawthorne.
\newblock {\em American Psychologist}, 73(4):308, 2018.

\bibitem{musick2021happens}
G.~Musick, T.~A. O'Neill, B.~G. Schelble, N.~J. McNeese, and J.~B. Henke.
\newblock What happens when humans believe their teammate is an ai? an investigation into humans teaming with autonomy.
\newblock {\em Computers in Human Behavior}, 122:106852, 2021.

\bibitem{o2018creating}
T.~A. O'Neill and E.~Salas.
\newblock Creating high performance teamwork in organizations.
\newblock {\em Human resource management review}, 28(4):325--331, 2018.

\bibitem{pontiggia2010network}
A.~Pontiggia and F.~Virili.
\newblock Network effects in technology acceptance: Laboratory experimental evidence.
\newblock {\em International Journal of Information Management}, 30(1):68--77, 2010.

\bibitem{salas2008teams}
E.~Salas, N.~J. Cooke, and M.~A. Rosen.
\newblock On teams, teamwork, and team performance: Discoveries and developments.
\newblock {\em Human factors}, 50(3):540--547, 2008.

\bibitem{schelble2023investigating}
B.~G. Schelble, C.~Flathmann, N.~J. McNeese, T.~O’Neill, R.~Pak, and M.~Namara.
\newblock Investigating the effects of perceived teammate artificiality on human performance and cognition.
\newblock {\em International Journal of Human--Computer Interaction}, 39(13):2686--2701, 2023.

\bibitem{westby2023collective}
S.~Westby and C.~Riedl.
\newblock Collective intelligence in human-ai teams: A bayesian theory of mind approach.
\newblock In {\em Proceedings of the AAAI Conference on Artificial Intelligence}, volume~37, pages 6119--6127, 2023.

\end{thebibliography}

\end{document}